\documentclass[aps,prx,twocolumn,groupedaddress,showkeys,showpacs,superscriptaddress,floatfix]{revtex4-2}

\usepackage{amsmath}
\usepackage{amssymb}
\usepackage{graphicx}
\usepackage{color}
\usepackage{orcidlink}

\begin{document}


\title{Heat-transfer fingerprint of Josephson breathers}



\author{Duilio De Santis\,\orcidlink{0000-0001-6501-9763}}
\email[]{duilio.desantis@unipa.it}
\affiliation{Dipartimento di Fisica e Chimica ``E.~Segr\`{e}", Group of Interdisciplinary Theoretical Physics, Università degli Studi di Palermo, I-90128 Palermo, Italy}

\author{Bernardo Spagnolo\,\orcidlink{0000-0002-6625-3989}}
\email[]{bernardo.spagnolo@unipa.it}
\affiliation{Dipartimento di Fisica e Chimica ``E.~Segr\`{e}", Group of Interdisciplinary Theoretical Physics, Università degli Studi di Palermo, I-90128 Palermo, Italy}
\affiliation{Radiophysics Department, Lobachevsky State University, 603950 Nizhniy Novgorod, Russia}

\author{Angelo Carollo\,\orcidlink{0000-0002-4402-2207}}
\email[]{angelo.carollo@unipa.it}
\affiliation{Dipartimento di Fisica e Chimica ``E.~Segr\`{e}", Group of Interdisciplinary Theoretical Physics, Università degli Studi di Palermo, I-90128 Palermo, Italy}

\author{Davide Valenti\,\orcidlink{0000-0001-5496-1518}}
\email[]{davide.valenti@unipa.it}
\affiliation{Dipartimento di Fisica e Chimica ``E.~Segr\`{e}", Group of Interdisciplinary Theoretical Physics, Università degli Studi di Palermo, I-90128 Palermo, Italy}

\author{Claudio Guarcello\,\orcidlink{0000-0002-3683-2509}} 
\email[]{cguarcello@unisa.it}
\affiliation{Dipartimento di Fisica ``E.~R.~Caianiello", Università degli Studi di Salerno, I-84084 Fisciano, Salerno, Italy}
\affiliation{INFN, Sezione di Napoli, Gruppo Collegato di Salerno - Complesso Universitario di Monte S. Angelo, I-80126 Napoli, Italy}


\date{\today}

\begin{abstract}
A sine-Gordon breather enhances the heat transfer in a thermally biased long Josephson junction. This solitonic channel allows for the tailoring of the local temperature throughout the system. Furthermore, the phenomenon implies a clear thermal fingerprint for the breather, and thus a `non-destructive' breather detection strategy is proposed here. Distinct breathing frequencies result in morphologically different local temperature peaks, which can be identified in an experiment.
\end{abstract}


\maketitle


\section{Introduction}

The soaring development of quantum technologies continuously propels the field of thermodynamics towards new fundamental and applied challenges, such as the accurate heat management at the nanoscale~\cite{Giazotto_2006,Pekola_2021}. This is the goal of \emph{caloritronics}~\cite{MartinezPerez_2014,Fornieri2017_2,Hwang_2020}, whose interest has been revived after the recent experimental demonstration of the phase-coherent control of the thermal transport in Josephson devices~\cite{Giazotto2012,Martinez-Perez_2014}. Superconducting phase coherence offers a unique knob for mastering heat flows. This feature has led researchers to conceive and implement nonlinear caloritronic devices for different applications, such as thermometry~\cite{Giazotto_2015,Zgirski_2018,Guarcello_2019} and refrigeration~\cite{Hofer_2016,Solinas_2016, Hwang_2023}, memories~\cite{Guarcello_2018_Memory,Ligato_2022} and engines~\cite{Scharf_2020,Cavaliere_2023}, routers~\cite{Timossi_2018,Hwang_2018,Acciai_2021} and switches~\cite{Sothmann_2017}, diffractors~\cite{Guarcello_2016_Diffr} and radiation detectors~\cite{Guarcello_2019_2,Paolucci_2023}. The strength of \emph{phase-coherent caloritronics} lies in the feasibility of adjusting the temperature by controlling the Josephson phase, e.g., via externally applied magnetic fields. 

Extended Josephson systems, such as long Josephson junctions (LJJs)~\cite{Barone_1982}, constitute an established research topic in applied superconductivity, thanks to their richness in terms of both physical phenomena and cutting-edge applications~\cite{Mazo_2014, Braginski_2019, Wustmann_2020, Wildermuth_2022, Lewis_2023}. From the viewpoint of thermal transport, the behavior of these devices is, however, largely unknown~\cite{Guarcello_2018, Guarcello_2018_1, Gua19}. It is then important to ask: can heat flows represent a new paradigm in such a context? Can this provide unprecedented means to experimentally investigate complex scenarios which have remained beyond reach for many years?

In this work, a fresh perspective is brought to the study of Josephson breathers, i.e., kink-antikink (or fluxon-antifluxon) oscillating bound states~\cite{Scott_2003, Dauxois_2006}. These excitations, which are notoriously hard to trace via standard techniques in the field of LJJs~\cite{Gulevich_2012, Monaco_2019, De_Santis_2022, De_Santis_2022_CNSNS, De_Santis_2022_NES, De_Santis_2023}, are shown to alter the heat transport in thermally biased junctions. The mastering of the local temperature within the system can thus be achieved via breathers, an intriguing fact which enhances their physical relevance. In addition, this unveiled property naturally sets the stage for a long sought non-destructive breather detection scheme, i.e., a protocol not involving the mode's breakup. Morphologically different local thermal profiles are found at distinct oscillation frequencies. The latter fact is useful in view of experiments and is a consequence of the analytical sine-Gordon (SG) breather waveform. By exploiting noise and ac driving for the excitation and the stabilization of the nonlinear breathing states~\cite{De_Santis_2023}, the robustness of the thermal fingerprint is demonstrated as well.

\section{The model}
\label{mod}

\begin{figure}[t!!]
\includegraphics[width=\columnwidth]{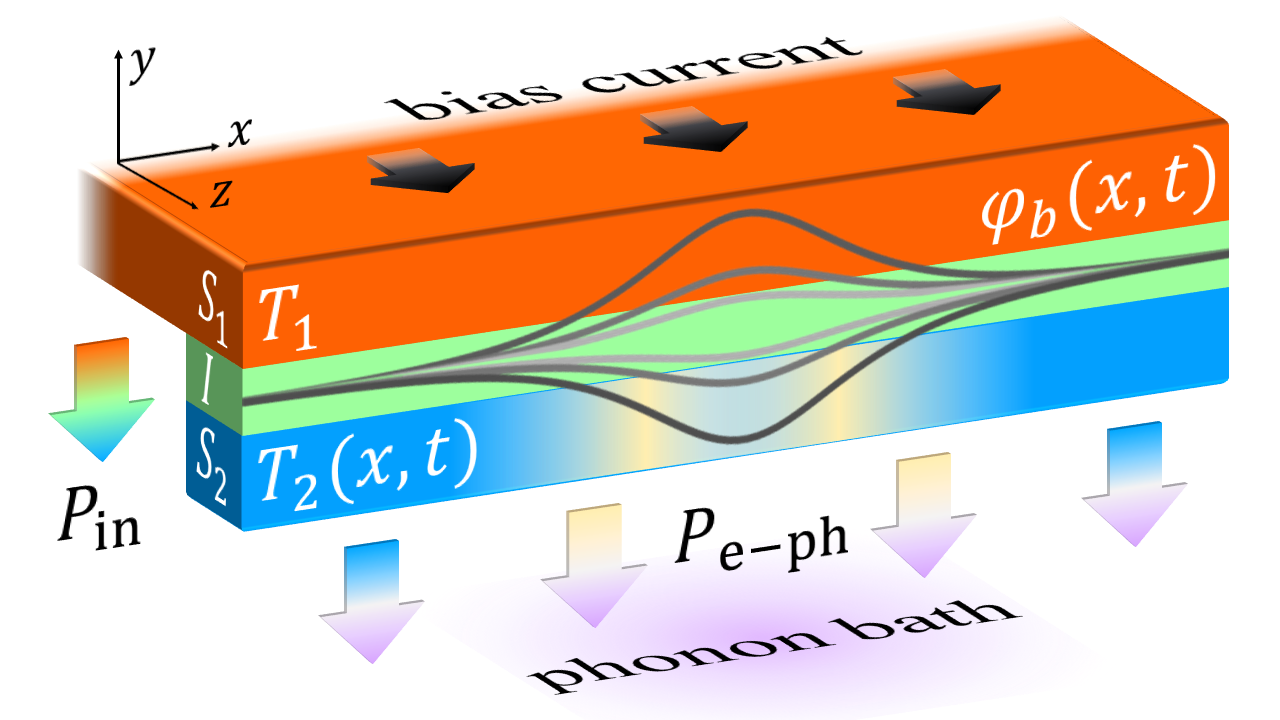}
\caption{Pictorial view of an overlap-geometry, current-driven LJJ, in the presence of a thermal bias. The temperature ${ T_1 }$ of the first electrode (${ S_1 }$) is fixed, whereas ${ S_2 }$ has a floating temperature ${ T_2 (x, t) }$. The latter electrode is also in thermal contact with a phonon bath at temperature ${ T_b }$. For ${ T_1 > T_2 (x, t) \geq T_b }$, the thermal power ${ P_{\mathrm{in}} (T_1, T_2, \varphi, V) }$ coming into ${ S_2 }$ is depicted, along with the outgoing term ${ P_{\mathrm{e-ph}} (T_2, T_b) }$, which is due to ${ S_2 }$'s quasiparticles coupling with the lattice phonons at ${ T_b }$. Lastly, the drawing illustrates the double-peaked local heating in ${ S_2 }$ caused by the breather oscillations, quantitatively discussed below.}
\label{fig:1}
\end{figure}

The dynamics of the LJJ is described via the perturbed SG equation for the Josephson phase ${ \varphi (\mathcal{X}, \mathcal{T}) }$~\cite{Barone_1982}
\begin{equation}
\label{eqn:1}
\frac{\partial^2 \varphi}{\partial \mathcal{X}^2} - \frac{\partial^2 \varphi}{\partial \mathcal{T}^2} - \alpha \frac{\partial \varphi}{\partial \mathcal{T}} = \sin \varphi - \eta \sin [\omega (\mathcal{T} - \mathcal{T}_0)] - \gamma_{n} (\mathcal{X}, \mathcal{T}) ,
\end{equation}
which is written in terms of dimensionless space and time variables, i.e., ${ \mathcal{X} = x / \lambda_J }$ and ${ \mathcal{T} = t \omega_p }$. Here, the characteristic scales are given by the Josephson penetration depth ${ \lambda_J = \sqrt{ \Phi_0 / \left( 2 \pi \mu_0 t_d J_c \right)} }$ and the Josephson plasma frequency ${ \omega_p = \sqrt{ 2 \pi J_c / \left( \Phi_0 C \right) } }$~\cite{Barone_1982}, where ${ t_d }$ is the effective magnetic thickness, ${ J_c = I_c / A }$ the Josephson critical current per unit area [with ${ A = L \times W }$ being the junction's area, and ${ L }$ (${ W }$) its length (width)], and ${ C }$ the junction's specific capacitance (${ \Phi_0 }$ and ${ \mu_0 }$ are the magnetic flux quantum and the vacuum permeability, respectively). Furthermore, in Eq.~\eqref{eqn:1}, ${ \alpha = 1 / \left( \omega_p R_a C \right) }$ is the damping coefficient (where ${ R_a }$ indicates the normal resistance per area), ${ \omega }$ (${ \eta }$) is the frequency (amplitude) of the external ac driving in units of ${ \omega_p }$ (${ I_c }$), ${ \mathcal{T}_0 }$ is a normalized time displacement, and ${ \gamma_{n} (\mathcal{X}, \mathcal{T}) }$ is a dimensionless noise current with zero average and autocorrelation function given by
\begin{equation}
\label{eqn:2}
\langle \gamma_{n}(\mathcal{X}_1, \mathcal{T}_1) \gamma_{n}(\mathcal{X}_2, \mathcal{T}_2) \rangle = 2 \alpha \Gamma \delta (\mathcal{X}_1 - \mathcal{X}_2) \delta (\mathcal{T}_1 - \mathcal{T}_2).
\end{equation}
The noise strength ${ \Gamma = 2 \pi \mathcal{L} k_B T / \left( \Phi_0 I_c \right) }$ is proportional to the normalized junction length ${ \mathcal{L} = L / \lambda_J }$ and the absolute temperature ${ T }$ ($ k_B $ is the Boltzmann constant). For an overlap-geometry LJJ, see Fig.~\ref{fig:1}, and null external magnetic fields, the imposed boundary conditions are ${ \left( \partial \varphi / \partial \mathcal{X} \right) \vert_{\mathcal{X} = 0, \mathcal{L}} = 0 }$.

Figure~\ref{fig:1} presents a sketch of the temperature-biased setup examined here, as well as the double-peaked local heating due to the breather oscillations. Specifically, the temperature ${ T_1 }$ of the first electrode (${ S_1 }$) is fixed, while ${ S_2 }$ has a floating temperature ${ T_2 (x, t) }$~\footnote[1]{This can be achieved via optimization of the electrodes' volumes~\cite{Guarcello_2018, Gua19}.}. The latter electrode is also in thermal contact with a phonon bath at known temperature ${ T_b }$, and the relation ${ T_1 > T_2 (x, t) \geq T_b }$ holds. The spatio-temporal behavior of the floating temperature is modeled by the diffusion equation~\cite{Guarcello_2018}
\begin{equation}
\label{eqn:3}
\frac{\partial}{\partial x} \left[ \kappa (T_2) \frac{\partial T_2}{\partial x} \right] + \mathcal{P}_{\mathrm{tot}}(T_1, T_2, T_b, \varphi, V) = c_v (T_2) \frac{\partial T_2}{\partial t} ,
\end{equation}
where ${ \kappa (T_2) }$ is the electronic heat conductivity~\cite{Fornieri_2017}, ${ \mathcal{P}_{\mathrm{tot}}(T_1, T_2, T_b, \varphi, V) = \mathcal{P}_{\mathrm{in}}(T_1, T_2, \varphi, V) - \mathcal{P}_{\mathrm{e-ph}}(T_2, T_b) }$, with ${ \mathcal{P}_{\mathrm{in}}(T_1, T_2, \varphi, V) }$ and ${ \mathcal{P}_{\mathrm{e-ph}}(T_2, T_b) }$ being, respectively, the ingoing and outgoing thermal power densities in ${ S_2 }$ [${ V \equiv V(x, t) }$ is the voltage drop]~\cite{Timofeev_2009}, see Fig.~\ref{fig:1}, and ${ c_v (T_2) }$ is the volume-specific heat capacity. Note that the LHS of Eq.~\eqref{eqn:3} describes the heat's spatial diffusion, while its RHS represents the variation of ${ S_2 }$'s internal energy.

The phase-dependent `in' thermal power density is structured as follows~\cite{Gua19}
\begin{equation}
\label{eqn:4}
\mathcal{P}_{\mathrm{in}}(T_1, T_2, \varphi, V) = \mathcal{P}_{\mathrm{qp}}(T_1, T_2, V) - \cos \varphi \; \mathcal{P}_{\mathrm{cos}}(T_1, T_2, V) .
\end{equation}
Here, the `qp' term is a quasiparticle contribution, i.e., it amounts to an ${ S_1 \; \mathrm{(hot)} \rightarrow S_2 \; \mathrm{(cold)} }$ incoherent energy flow, whereas the `cos' term stems from the energy-carrying tunneling events involving the destruction and recombination of Cooper pairs in both ${ S_1 }$ and ${ S_2 }$~\cite{Maki_1965, Golubev_2013}.

The initial condition for Eq.~\eqref{eqn:3} is ${ T_2 \vert_{t = 0} = T_b }$, and the edges of the device are assumed to be thermally isolated, implying ${ \left( \partial T_2 / \partial x \right) \vert_{x = 0, L} = 0 }$.

Detailed information regarding all the above expressions, including their physical significance and the numerical means to handle them, is given in Apps.~\ref{appA} and \ref{appB}. In what follows, an LJJ composed by Nb/${ \mathrm{AlO}_x }$/Nb is considered, with ${ L = 300 \; \mu \mathrm{m} }$, ${ W = 0.5 \; \mu \mathrm{m} }$, ${ d_2 = 0.1 \; \mu \mathrm{m} }$ (thickness of ${ S_2 }$), ${ d = 1 \; \mathrm{nm} }$ (thickness of the insulating layer), ${ R_a = 50 \; \Omega \; \mu \mathrm{m}^2 }$, and ${ C = 100 \; \mathrm{fF} \; \mu \mathrm{m}^{-2} }$. Such a device is thermally biased by setting ${ T_1 = 7 \; \mathrm{K} }$ and ${ T_b = 4.2 \; \mathrm{K} }$~\cite{Guarcello_2018}. As shown in App.~\ref{appC}, by accounting for the temperature dependence in both ${ t_d (T_1, T_2) }$ and ${ I_c (T_1, T_2) }$, one can estimate the values~\footnote[3]{The approximation ${ T_2 = T_b }$ is made here, since ${ T_2 }$'s variations (discussed later in the work) are negligible for the sake of these calculations.} ${ \lambda_J \approx 8 \; \mu \mathrm{m} }$ (thus ${ \mathcal{L} = L / \lambda_J \approx 37 }$), ${ \omega_p \approx 0.95 \; \mathrm{THz} }$, ${ \alpha \approx 0.2 }$, and ${ \Gamma \approx 0.0026 }$~\footnote[4]{The latter quantity explicitly depends on the temperature ${ T }$, see Eq.~\eqref{eqn:2}. To display the results' robustness even in the `worst-case' noise scenario, the highest temperature within the system, i.e., ${ T = T_1 }$, is taken.}. Before proceeding, it should be also stressed that the overall features found below apply to a wide range of parameter values. The currently chosen set is just meant to provide a realistic example.

\section{Breather-enhanced thermal transport}

The effects of breathers on the evolution of the temperature ${ T_2(\mathcal{X}, \mathcal{T}) }$ are studied in the following way. First, the electrode ${ S_2 }$ is allowed to fully thermalize in the absence of excitations, e.g., by taking ${ \varphi (\mathcal{X}, \mathcal{T} < \mathcal{T}_0) }$ in an unperturbed scenario. Equation~\eqref{eqn:1} is then solved for ${ \mathcal{T}_0 \leq \mathcal{T} \leq \mathcal{T}_f }$, and the generation of a stabilized breather centered at ${ \mathcal{X} = \mathcal{X}_0 }$ is mimicked by imposing ${ \varphi \vert_{\mathcal{T} = \mathcal{T}_0} = \varphi^0 \vert_{\mathcal{T} = \mathcal{T}_0} }$ and ${ \left( \partial \varphi / \partial \mathcal{T} \right) \vert_{\mathcal{T} = \mathcal{T}_0} = \left( \partial \varphi^0 / \partial \mathcal{T} \right) \vert_{\mathcal{T} = \mathcal{T}_0} }$, with ${ \varphi^0 (\mathcal{X}, \mathcal{T}) = \varphi_b (\mathcal{X}, \mathcal{T}) + \varphi_v (\mathcal{T}) }$ being the `breather plus vacuum' state~\cite{Lomdahl_1986}
\begin{align}
\label{eqn:5}
\nonumber \varphi^0 (\mathcal{X}, \mathcal{T}) &= 4 \; \mathrm{atan} \left\lbrace \frac{\sqrt{1 - \omega^2}}{\omega} \frac{\cos \left[ \omega \left( \mathcal{T} - \mathcal{T}_0 \right) + \pi - 2 \theta \right]}{\cosh \left[ \sqrt{1 - \omega^2} \left( \mathcal{X} - \mathcal{X}_0 \right) \right]} \right\rbrace \\
& + \frac{\eta \sin \left[ \omega \left( \mathcal{T} - \mathcal{T}_0 \right) - \theta \right]}{\sqrt{(1 - \omega^2)^2 + \alpha^2 \omega^2}} , \; \; \; \tan \theta = \frac{\alpha \omega}{1 - \omega^2} ,
\end{align}
which is known to lock to an ac force of suitable amplitude, i.e., for ${ \eta \approx \eta_{\mathrm{th}} (\omega) = \frac{2 \alpha (1 - \omega^2) \; \mathrm{asin} \sqrt{1 - \omega^2}}{K(1 - \omega^2) - E(1 - \omega^2)} }$ (${ K }$ and ${ E }$ are, respectively, complete elliptic integrals of the first and the second kind)~\cite{Lomdahl_1986}. The choices ${ x_0 = \mathcal{X}_0 \lambda_J = 150 \; \mu \mathrm{m} }$, ${ t_0 = \mathcal{T}_0 / \omega_p = 2 \; \mathrm{ns} }$, and ${ t_f = \mathcal{T}_f / \omega_p = 10 \; \mathrm{ns} }$ are made. It is worth adding that the below long-time temperature behaviors do not depend on the value of the excitation-free thermalization time ${ t_0 }$.

\begin{figure}[t!!]
\includegraphics[width=\columnwidth]{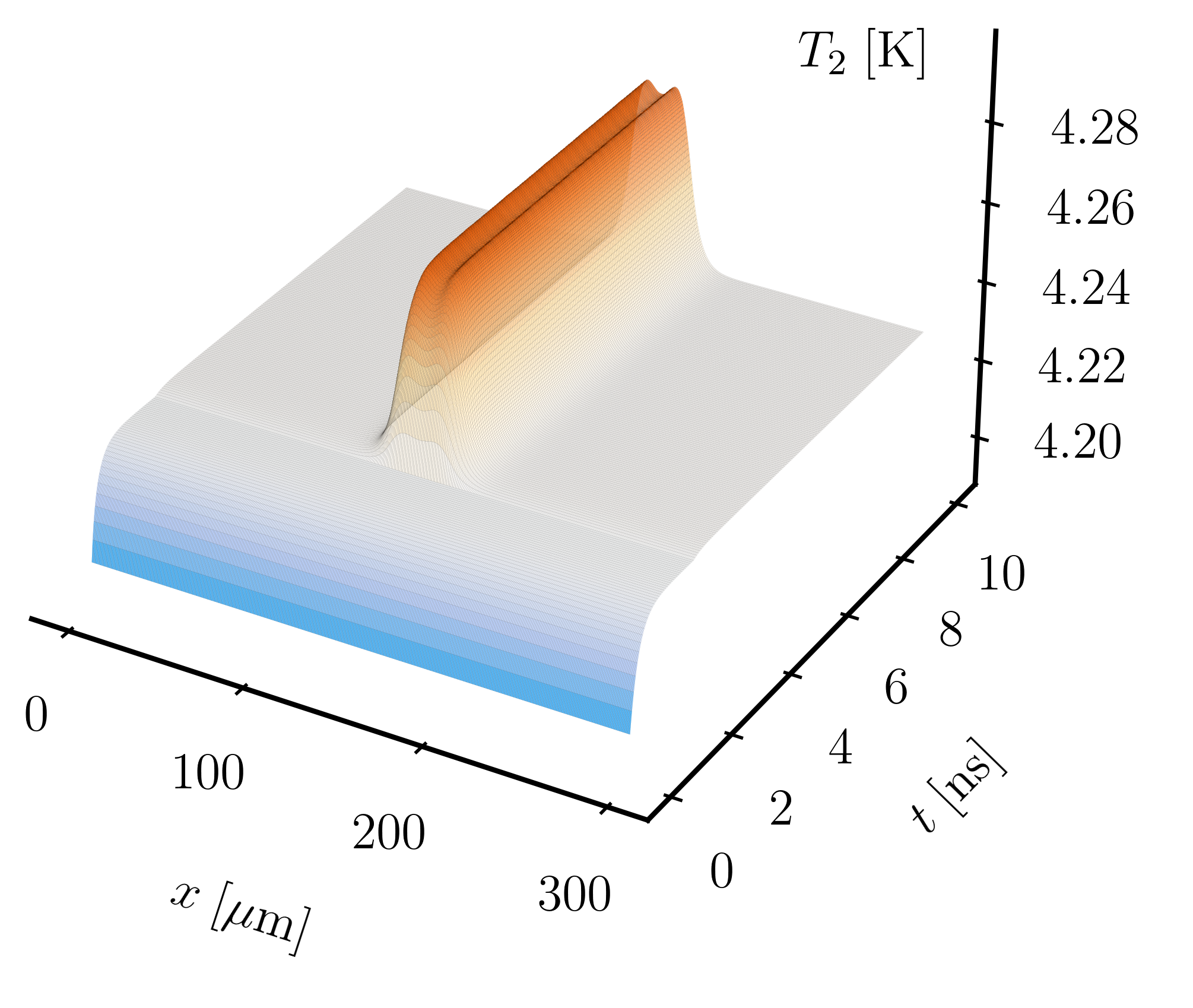}
\caption{Spatio-temporal view of the temperature ${ T_2 }$. The generation of a stabilized breather centered at ${ x_0 = 150 \; \mu \mathrm{m} }$ is mimicked at ${ t_0 = 2 \; \mathrm{ns} }$, that is, after ${ S_2 }$ has thermalized at the steady value of ${ \sim 4.23 \; \mathrm{K} }$. The plot displays the double-peaked local heating, with ${ \max \left\lbrace T_2 \right\rbrace \gtrsim 4.27 \; \mathrm{K} }$, caused by the breather. Parameter values: ${ \omega = 0.5 }$, ${ \eta = 0.33 }$, and ${ \Gamma = 0 }$.}
\label{fig:2}
\end{figure}
Setting ${ \Gamma = 0 }$, a typical simulation outcome for ${ T_2 (x, t) }$ is shown in Fig.~\ref{fig:2} for ${ \omega = 0.5 }$ and ${ \eta = 0.33 \approx \eta_{\mathrm{th}} (\omega = 0.5) }$. One can first appreciate the temperature's relaxation towards the (excitation-free) steady value of ${ \sim 4.23 \; \mathrm{K} }$ within ${ t = t_0 }$. Strikingly, for ${ t > t_0 }$, a local exponential growth of the temperature is observed in correspondence to the breather, leading to a persisting double-peaked local heating, with ${ \max \left\lbrace T_2 \right\rbrace \gtrsim 4.27 \; \mathrm{K} }$. Such a profile is a natural consequence of the breather being a two-soliton bound state. A clear thermal fingerprint for the breather is thus brought to light. The latter mode is indeed notoriously hard to track via mean voltage measurements due to its fast (zero-averaging) oscillations, and the few proposals currently available in the literature resort to the breather's destruction for probing purposes~\cite{Gulevich_2012, De_Santis_2022, De_Santis_2023}. By virtue of the characteristic `${ \cos \varphi }$' dependence in Eq.~\eqref{eqn:4}, the breather's influence within the thermal realm is nonvanishing, and the detection does not require its breakup{\textemdash}a fact which makes this framework particularly appealing.

\begin{figure}[t!!]
\includegraphics[width=\columnwidth]{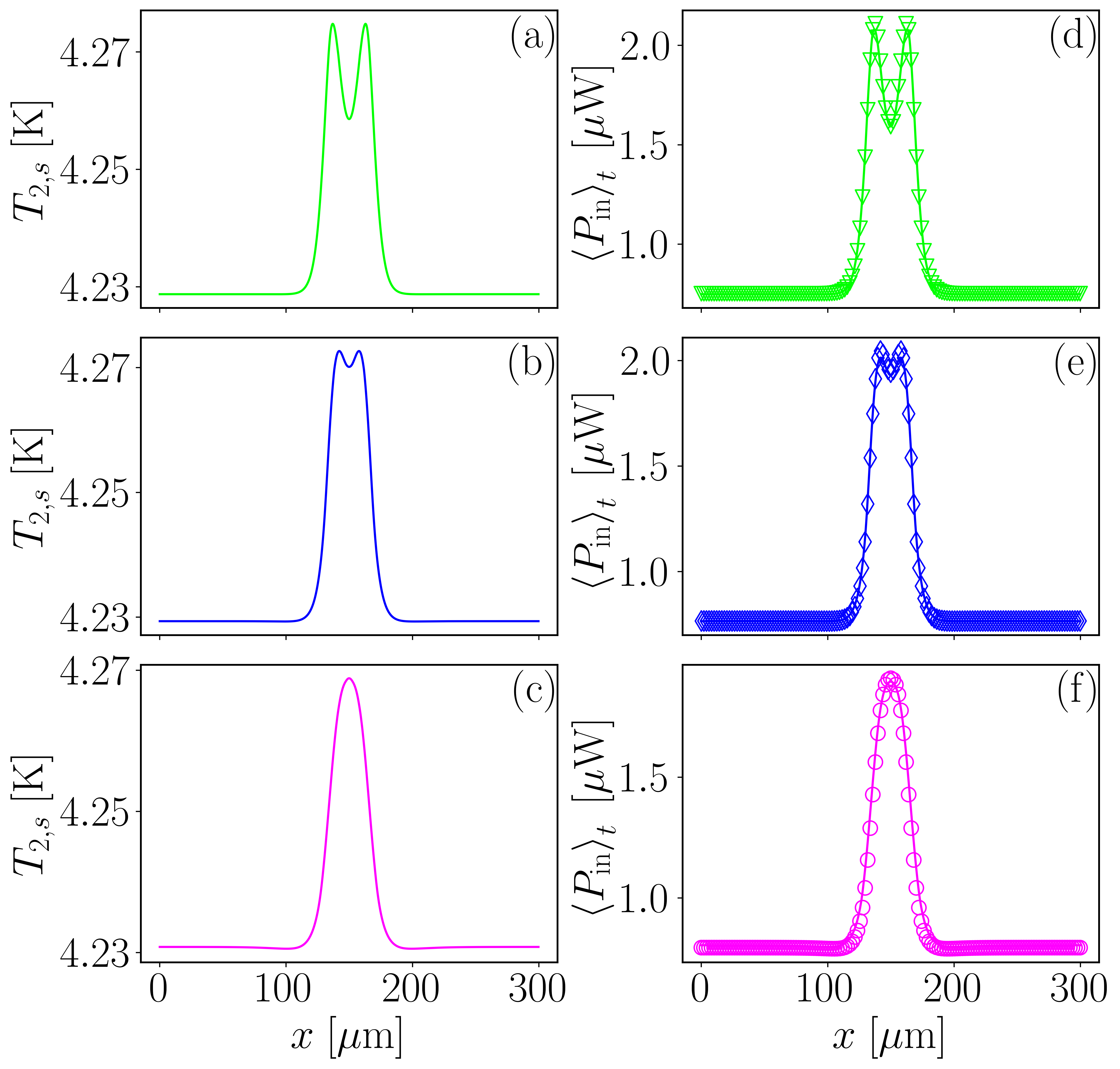}
\caption{Panels~(a)-(c): Stationary temperature curves (${ T_{2,s} }$) at different breather frequencies. Panels~(d)-(f): Time-averaged thermal powers ${ \left\langle P_{\mathrm{in}} \right\rangle_t }$ corresponding to the simulations from panels~(a)-(c) (lines) and their analytical counterparts, i.e., ${ \left\langle P_{\mathrm{in}} (T_1, T_b, \varphi^0, V^0) \right\rangle_t }$ (triangles, diamonds, and circles). In the panels, ${ \Gamma = 0 }$ is set, and the green, blue, and purple colors stand for the combinations ${ \omega = 0.3 }$ and ${ \eta = 0.3 }$, ${ \omega = 0.5 }$ and ${ \eta = 0.33 }$, and ${ \omega = 0.7 }$ and ${ \eta = 0.31 }$, respectively.}
\label{fig:3}
\end{figure}
It follows from the analytical SG breather solution, see Eq.~\eqref{eqn:5}, that the mode's morphology, e.g., its oscillation amplitude and width, encodes information on the breathing frequency~\cite{Scott_2003, Dauxois_2006}. It is hence reasonable to expect differently shaped thermal profiles for distinct frequencies, with lower ${ \omega }$ values yielding more pronounced double-peaked patterns (since the corresponding breathers are closer to unbound kink-antikink pairs). This is confirmed in Fig.~\ref{fig:3}(a)-(c), which displays three stationary temperature curves (${ T_{2,s} }$) at different breather frequencies. Here, the green, blue, and purple lines are obtained, respectively, for ${ \omega = \left\lbrace 0.3, 0.5, 0.7 \right\rbrace }$ and ${ \eta = \left\lbrace 0.3, 0.33, 0.31 \right\rbrace }$, where ${ \eta \approx \eta_{\mathrm{th}} (\omega) }$. Note that in the low-frequency representative case a markedly double-peaked waveform emerges, with ${ \max \left\lbrace T_{2,s} \right\rbrace \gtrsim 4.27 \; \mathrm{K} }$, while in the high-frequency one the resulting curve is essentially bell-shaped, with ${ \max \left\lbrace T_{2,s} \right\rbrace \lesssim 4.27 \; \mathrm{K} }$. The result at ${ \omega = 0.5 }$ is somewhat in between the previous two, as one may guess. Far from electrode's center, ${ T_{2,s} \sim 4.23 \; \mathrm{K} }$ is obtained in all scenarios.

As hinted above, the temperature's envelope reflects the behavior of the thermal power ${ P_{\mathrm{in}} }$. This structural analogy is clearly demonstrated through Fig.~\ref{fig:3}(d)-(f), where the lines indicate the time-averaged thermal powers ${ \left\langle P_{\mathrm{in}} \right\rangle_t }$~\footnote[5]{The time average is intended for ${ t \geq t_0 }$, with the simulations' being performed in analogous fashion to that presented in Fig.~\ref{fig:2}.} relative to the simulations of Eq.~\eqref{eqn:1} and Eq.~\eqref{eqn:3} discussed in panels~(a)-(c). In particular, the local enhancement of the average thermal power is higher for the prominent double-peaked (green) profile, for which ${ \max \left\lbrace \left\langle P_{\mathrm{in}} \right\rangle_t \right\rbrace \gtrsim 2 \; \mu \mathrm{W} }$, whereas the bell-shaped (purple) one is characterized by ${ \max \left\lbrace \left\langle P_{\mathrm{in}} \right\rangle_t \right\rbrace \lesssim 2 \; \mu \mathrm{W} }$. The ${ \omega = 0.5 }$ (blue) outcome falls once again in between the prior two. In all cases, ${ \left\langle P_{\mathrm{in}} \right\rangle_t \lesssim 1 \; \mu \mathrm{W} }$ is observed away from the breather.

Figure~\ref{fig:3}(d)-(f) features also a comparison between the above mentioned ${ \left\langle P_{\mathrm{in}} \right\rangle_t }$ curves and their analytical counterparts. More specifically, the green triangles (${ \omega = 0.3 }$), the blue diamonds (${ \omega = 0.5 }$), and the purple circles (${ \omega = 0.7 }$) are calculated directly from Eq.~\eqref{eqn:5} as ${ \left\langle P_{\mathrm{in}} (T_1, T_b, \varphi^0, V^0) \right\rangle_t }$, with ${ V^0 = [\Phi_0 / (2 \pi)] \left( \partial \varphi^0 / \partial t \right) }$ being the voltage profile associated to ${ \varphi^0 }$, without using Eqs.~\eqref{eqn:1} and \eqref{eqn:3}. The great accordance displayed is yet another indication that the unveiled phenomenology can be rigorously ascribed to the breather waveform.

\begin{figure}[t!!]
\includegraphics[width=\columnwidth]{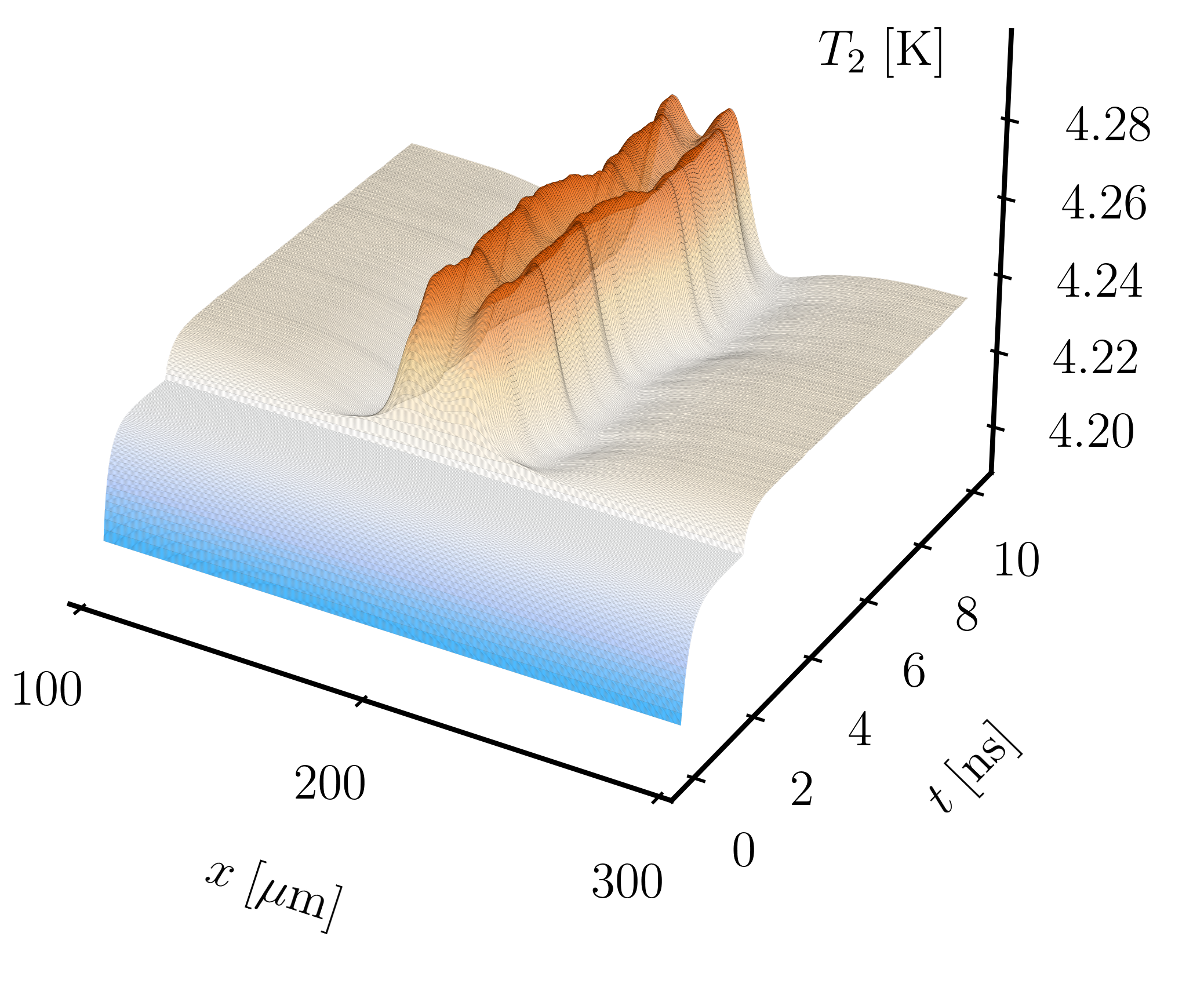}
\caption{Spatio-temporal view of the temperature ${ T_2 }$. A noise-induced and ac-locked breather, centered roughly at ${ 200 \; \mu \mathrm{m} }$, is observed for ${ t \gtrsim t_0 = 2 \; \mathrm{ns} }$, i.e., after ${ S_2 }$'s thermalization at the steady value of ${ \sim 4.23 \; \mathrm{K} }$ has occurred. Focusing on the region ${ [100, 300] \; \mu \mathrm{m} }$, the breather is shown to induce a double-peaked local heating, with ${ \max \left\lbrace T_2 \right\rbrace \gtrsim 4.27 \; \mathrm{K} }$. Parameter values: ${ \omega = 0.6 }$, ${ \eta = 0.59 }$, and ${ \Gamma = 0.0026 }$.}
\label{fig:4}
\end{figure}

\section{Noise-induced breathers}

Pairing the revealed thermal fingerprint with a reliable breather excitation and stabilization technique is an important task in view of an experimental detection. To this end, it has been recently established that the combination of noise and ac forcing can lead to the emergence of long-time stable breather modes in random locations~\cite{De_Santis_2023}. 

In the simulation shown in Fig.~\ref{fig:4}, the noisy junction (${ \Gamma = 0.0026 }$, see Sec.~\ref{mod}) is driven with the frequency/amplitude combination ${ \omega = 0.6 }$ and ${ \eta = 0.59 }$~\cite{De_Santis_2023}. After the excitation-free thermalization of the electrode ${ S_2 }$ for ${ t < t_0 }$, a noise-induced and ac-locked breather arises close to ${ 200 \; \mu \mathrm{m} }$, which motivates the plot's focus on the region ${ [100, 300] \; \mu \mathrm{m} }$. Interestingly, the distinctive traits of the breather's influence on the temperature ${ T_2 }$ are preserved even in the stochastic scenario. As one expects, now the position of the solitonic state's center slightly fluctuates in time, along with its amplitude, but the overall behavior is analogous to that presented in Fig.~\ref{fig:2}. A double-peaked local heating, with ${ \max \left\lbrace T_2 \right\rbrace \gtrsim 4.27 \; \mathrm{K} }$, is indeed clearly seen. In short, noisy and ac-driven LJJs represent natural candidates for unveiling the breather's thermal fingerprint, which is non-destructive.

\section{Conclusions}

It is demonstrated that SG breathers enhance the heat transfer in thermally biased, overlap-geometry LJJs. This brings to light a soliton-based mechanism for mastering the local temperature within the system. Another important point is that this effect allows for the design of a non-destructive breather detection strategy. Notably, distinct breathing frequencies yield morphologically different local temperature peaks, which are well understood in terms of the analytical SG breather profile and can be experimentally identified. To illustrate the robustness of the results, this work pairs the above thermal fingerprint with a reliable breather and stabilization technique, via the combined action of noise and ac driving~\cite{De_Santis_2023}.

The developed ideas are expected to find application even beyond the LJJ device. In particular, it seems reasonable to look at discrete systems, and exploit a similar approach to study and probe, for example, the elusive oscillobreather states in parallel arrays of thermally biased superconducting junctions~\cite{Mazo_2003}. Furthermore, a connection between the present scenario and the more general context of soliton-sustained heat propagation in various devices, such as wires and nanotubes~\cite{Sciacca_2020,Sciacca_2022}, naturally comes to mind.

\appendix
\section{Modelling of the thermally biased long Josephson junction}
\label{appA}

To model the evolution of the floating temperature, one should discuss the typical length scale for thermalization within the electrode. In the diffusive regime, the inelastic scattering length can be considered ${ l_{\mathrm{in}} = \sqrt{\tau_s \mathcal{D}} \approx 0.3 \; \mu \mathrm{m} }$ for Nb at ${ 4.2 \; \mathrm{K} }$, where ${ \tau_s }$ is the quasiparticle scattering lifetime and ${ \mathcal{D} = \sigma_N / \left( e^2 N_F \right) }$ is the diffusion constant, with ${ \sigma_N }$ and ${ N_F }$ being, respectively, the electrical conductivity in the normal state and the density of states at the Fermi energy (${ e }$ is the electron charge). Since exclusively the junction length ${ L }$ is much larger than ${ l_{\mathrm{in}} }$, ${ S_2 }$ essentially behaves as a 1-d diffusive superconductor, and Eq.~\eqref{eqn:3} holds~\cite{Guarcello_2018}. Furthermore, in Eq.~\eqref{eqn:3}, the electronic heat conductivity ${ \kappa (T_2) }$ reads~\cite{Fornieri_2017}
\begin{equation}
\label{eqn:S2}
\begin{aligned}
\kappa (T_2) &= \frac{\sigma_N}{2 e^2 k_B T_2^2} \int_{- \infty}^{+ \infty} \varepsilon^2 \\
& \times \frac{\cos^2 \left\lbrace \mathrm{Im} \left[ \mathrm{arctanh} \left( \frac{\Delta(T_2)}{\varepsilon + i \gamma} \right) \right] \right\rbrace}{\cosh^2 \left( \frac{\varepsilon}{2 k_B T_2} \right)} d \varepsilon .
\end{aligned}
\end{equation}
Here, the BCS-like superconducting gap ${ \Delta(T) = \Delta \tanh \left( 1.74 \sqrt{ T_c / T - 1 } \right) }$ is employed~\cite{Senapati_2011}, with ${ \Delta = 1.764 k_B T_c }$, ${ T_c = 9.2 \; \mathrm{K} }$ being the critical temperature for Nb, and ${ \gamma = 10^{-4} \Delta }$ the Dynes broadening parameter~\cite{Dynes_1978}.

In the adiabatic limit, i.e., when ${ e V \ll \min \left\lbrace k_B T_1 , k_B T_2 , \Delta(T_1) , \Delta(T_2) \right\rbrace }$, one can write~\cite{Golubev_2013}
\begin{equation}
\label{eqn:S4}
\begin{aligned}
\mathcal{P}_{\mathrm{qp}}(T_1, T_2, V) &= \frac{1}{e^2 R_a d_2} \int_{- \infty}^{+ \infty} \mathcal{N} (\varepsilon - e V, T_1) \\ & \times \mathcal{N}(\varepsilon, T_2) \left( \varepsilon - e V \right) \\ & \times \left[ f(\varepsilon - e V, T_1) - f(\varepsilon, T_2)\right] d \varepsilon
\end{aligned}
\end{equation}
and
\begin{equation}
\label{eqn:S5}
\begin{aligned}
\mathcal{P}_{\mathrm{cos}}(T_1, T_2, V) &= \frac{1}{e^2 R_a d_2} \int_{- \infty}^{+ \infty} \mathcal{N} (\varepsilon - e V, T_1) \\ & \times \mathcal{N}(\varepsilon, T_2) \frac{\Delta(T_1) \Delta(T_2)}{\varepsilon} \\ & \times \left[ f(\varepsilon - e V, T_1) - f(\varepsilon, T_2)\right] d \varepsilon ,
\end{aligned}
\end{equation}
in which ${ \mathcal{N}(\varepsilon, T) = \left\vert \mathrm{Re} \left[ \frac{ \varepsilon + i \gamma }{ \sqrt{ \left( \varepsilon + i \gamma \right)^2 - \Delta(T)^2 } } \right] \right\vert }$ is the reduced superconducting density of state, and ${ f (\varepsilon, T) = \left[ 1 + e^{ \varepsilon / \left( k_B T \right) } \right]^{-1} }$ is the Fermi distribution function. The breather oscillations quickly average to zero voltage, i.e., ${ \left\langle V_b \right\rangle_t \approx 0 }$ is observed over times much smaller than the characteristic time of Eq.~\eqref{eqn:3}, thus satisfying the above adiabatic condition. Furthermore, it should be emphasized that the essence of the phenomenology, that is, the thermal profiles discussed in the main text (Figs.~\ref{fig:2}-\ref{fig:4}), lies in the cosine ${ \varphi }$-dependence alone.

The `e-ph' thermal power density accounts for the energy exchange between electrons and phonons in the superconductor, and it is given by~\cite{Timofeev_2009}
\begin{equation}
\begin{aligned}
\label{eqn:S6}
\mathcal{P}_{\mathrm{e-ph}} &= \frac{- \Sigma}{96 \zeta (5) k_B^5} \int_{- \infty}^{+ \infty} E d E \int_{- \infty}^{+ \infty} \varepsilon^2 \mathrm{sign}(\varepsilon) M_{E, E + \varepsilon} \\
& \times \left\lbrace \mathrm{coth} \left( \frac{\varepsilon}{2 k_B T_b} \right) \left[ \mathcal{F}(E, T_2) - \mathcal{F}(E + \varepsilon, T_2) \right] \right. \\ & \left. \vphantom{\mathrm{coth} \left( \frac{\varepsilon}{2 k_B T_b} \right)} - \mathcal{F}(E, T_2) \mathcal{F}(E + \varepsilon, T_2) + 1 \right\rbrace d \varepsilon ,
\end{aligned}
\end{equation}
\sloppy where ${ \Sigma }$ is the electron-phonon coupling constant, ${ \zeta }$ is the Riemann zeta function, ${ M_{E,E'} = \mathcal{N}(E, T_2) \mathcal{N}(E', T_2) \left[ 1 - \Delta(T_2)^2 / \left( E E' \right) \right] }$, \nolinebreak and ${ \mathcal{F} (\varepsilon, T_2) = \tanh \left[ \varepsilon / \left( 2 k_B T_2 \right) \right] }$. Here, the lattice phonons of the superconductor are assumed to be thermalized with the substrate residing at ${ T_b }$ by virtue of the vanishing Kapitza resistance between the thin metallic films and the substrate at low temperatures~\cite{Giazotto_2006}.

The RHS of Eq.~\eqref{eqn:3} involves ${ c_v (T_2) = T_2 \left( d \mathcal{S} / d T_2 \right) }$, which is defined in terms of the electronic entropy density of ${ S_2 }$~\cite{Solinas_2016}
\begin{equation}
\begin{aligned}
\label{eqn:S7}
\mathcal{S}(T_2) &= - 4 k_B N_F \int_{0}^{+ \infty} \mathcal{N}(\varepsilon, T_2) \left\lbrace \left[ 1 - f(\varepsilon, T_2) \right] \right. \\
& \left. \times \mathrm{ln} \left[ 1 - f(\varepsilon, T_2) \right] + f(\varepsilon, T_2) \mathrm{ln} f(\varepsilon, T_2) \right\rbrace d \varepsilon .
\end{aligned}
\end{equation}

The following parameter values are considered: ${ \sigma_N = 6.7 \times 10^6 \; \Omega^{-1} \; \mathrm{m}^{-1} }$, ${ N_F = 10^{47} \; \mathrm{J}^{-1} \; \mathrm{m}^{-3} }$, and ${ \Sigma = 3 \times 10^9 \; \mathrm{W} \; \mathrm{m}^{-3} \; \mathrm{K}^{-5} }$.

To conclude, it is perhaps worth mentioning that the present framework can be rephrased to apply for junctions involving non-identical electrodes as well.

\section{Numerical details}
\label{appB}

\begin{figure*}[t!!]
\includegraphics[width=\textwidth]{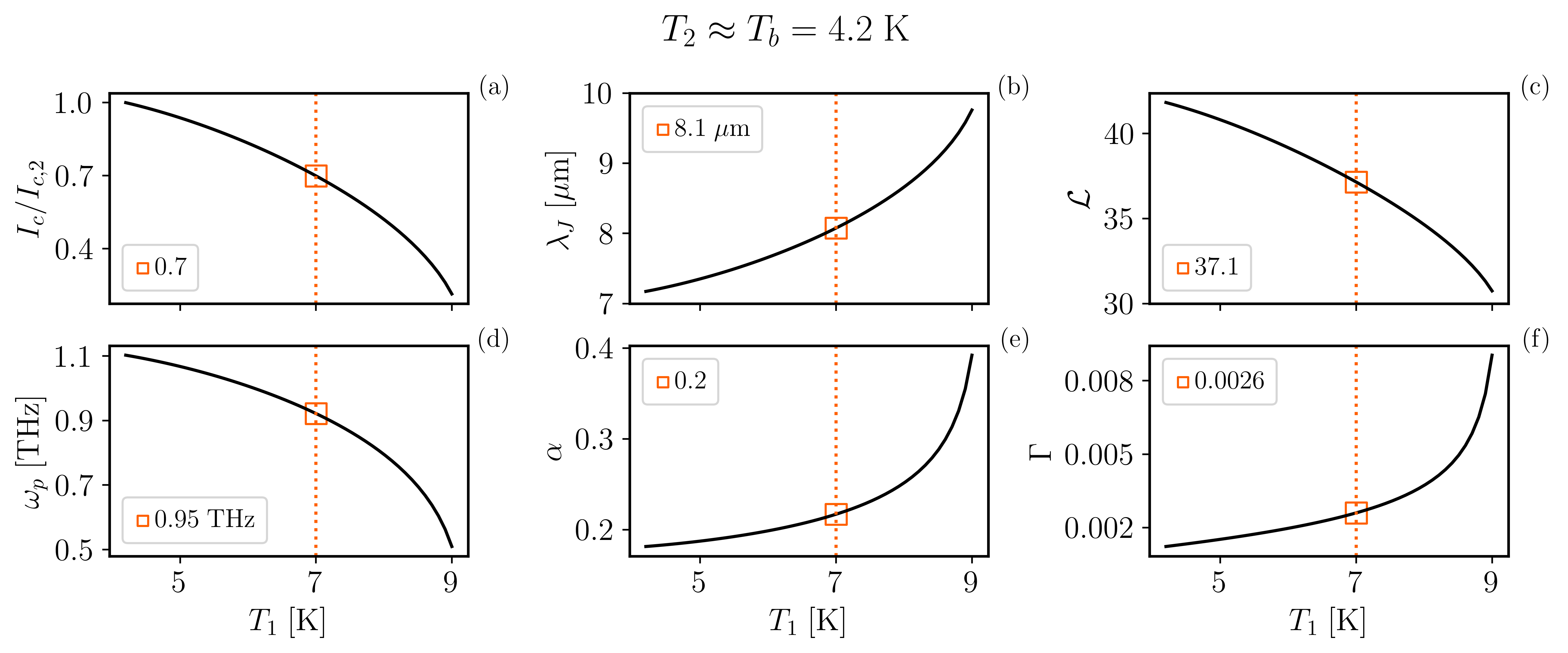}
\caption{Josephson critical current ${ I_c }$ in units of ${ I_{c, 2} = I_c (T_2, T_2) }$~[panel~(a)], Josephson penetration depth~${ \lambda_J }$~[panel~(b)], normalized junction length ${ \mathcal{L} }$~[panel~(c)], Josephson plasma frequency ${ \omega_p }$~[panel~(d)], dissipation coefficient ${ \alpha }$~[panel~(e)], and noise strength ${  \Gamma }$~[panel~(f)] as a function of the temperature ${ T_1 }$. Here, the approximation ${ T_2 = T_b = 4.2 \; \mathrm{K} }$ is used, and the orange squares denote the parameter values at ${ T_1 = 7 \; \mathrm{K} }$, i.e., those employed throughout the work.}
\label{fig:S1}
\end{figure*}
Implicit finite-difference schemes are used to solve both the perturbed SG equation and the diffusion equation, keeping the same spatio-temporal grid in the two cases. In particular, the spatial domain is divided into $ \mathcal{N} $ cells of length ${ \Delta x }$ and the temporal domain into $ \mathcal{M} $ intervals of duration ${ \Delta t }$. The perturbed SG equation is handled in the same way as in Ref.~\cite{De_Santis_2023}, see its Supplemental Material for a complete account of this matter. In Eq.~\eqref{eqn:3}, indicating the space-time restriction of ${ T_2 (x, t) }$ as ${ {(T_2)}_n^m = T_2 (n \; \Delta x, m \; \Delta t) }$, for ${ n = 1, ..., \mathcal{N} }$ and ${ m = 1, ..., \mathcal{M} }$, the derivatives are~\cite{Ames_1977}
\begin{equation}
\label{eqn:S8}
\begin{aligned}
& \frac{\partial T_2}{\partial x} \approx \frac{1}{2 \Delta x} \left[ {(T_2)}_{n + 1}^{m} - {(T_2)}_{n - 1}^{m} \right] , \\
& \frac{\partial T_2}{\partial t} \approx \frac{1}{\Delta t} \left[ {(T_2)}_{n}^{m + 1} - {(T_2)}_{n}^{m} \right] , \\
& \frac{\partial^2 T_2}{\partial x^2} \approx \frac{1}{2 \Delta x^2} \left[ {(T_2)}_{n + 1}^{m + 1} - 2 {(T_2)}_{n}^{m + 1} + {(T_2)}_{n - 1}^{m + 1} \right. \\ & \left. \hphantom{\frac{\partial^2 T_2}{\partial x^2}} + {(T_2)}_{n + 1}^{m} - 2 {(T_2)}_{n}^{m} + {(T_2)}_{n - 1}^{m} \right] .
\end{aligned}
\end{equation}
By also applying both the initial and the boundary conditions, a tridiagonal system of equations is obtained. The latter's resolution, achievable through, e.g., the Thomas' algorithm~\cite{Press_1992}, determines the unknown values ${ {(T_2)}_{n}^{m + 1} }$, given the previous ones ${ {(T_2)}_{n}^{m} }$, with ${ n = 1, ..., \mathcal{N} }$. Throughout the work, the values of the discretization steps are ${ \Delta x = 0.4 \; \mu \mathrm{m} }$ and ${ \Delta t = 0.01 \; \mathrm{ps} }$ (i.e., ${ \Delta \mathcal{X} = \Delta x / \lambda_J = 0.05 }$ and ${ \Delta \mathcal{T} = \Delta t \; \omega_p = 0.01 }$).

There is one more technical aspect concerning the solution of Eq.~\eqref{eqn:3}, that is, the ${ T_2 }$-dependent integrals in Eqs.~\eqref{eqn:S2}-\eqref{eqn:S7}. At each time step of the implicit scheme, in principle, one would need to numerically compute the latter objects by using the instantaneous ${ T_2 }$ temperature profile [note that Eqs.~\eqref{eqn:S4}-\eqref{eqn:S5} also depend on the instantaneous voltage]. Given the time-consuming nature of this task, another approach is pursued here. Via preliminary testing, very refined grids covering the entire variation range of the required quantities, i.e., ${ T_2 }$ and ${ V }$, are constructed to evaluate all the integrals beforehand. These results are stored, and subsequently (if needed) standard interpolation routines are called to fill the gaps. The use of \texttt{scipy}'s \texttt{integrate.quad} (which is based on the \texttt{Fortran} library \texttt{QUADPACK}), as well as its interpolation schedules, is acknowledged.

\section{Sine-Gordon model in the thermally biased scenario: parameter values}
\label{appC}

The parameters introduced within the SG framework are influenced by the temperatures ${ T_1 }$ and ${ T_2 }$, and therefore they must be set properly. In particular, the effective magnetic thickness is given by~\cite{Martinez-Perez_2014}
\begin{equation}
\begin{aligned}
\label{eqn:S9}
t_d (T_1, T_2) &= \lambda_L (T_1) \tanh \left[ \frac{d_1}{2 \lambda_L (T_1)} \right] \\ & + \lambda_L (T_2) \tanh \left[ \frac{d_2}{2 \lambda_L (T_2)} \right] + d ,
\end{aligned}
\end{equation}
where ${ \lambda_L (T) = \frac{\lambda_L^0}{\sqrt{1 - (T / T_c)^4}} }$ is the London penetration depth, and ${ d_i }$ is the thickness of the electrode ${ S_i }$. The above assumption of a fixed temperature ${ T_1 }$ amounts to consider a very large ${ S_1 }$ volume, thus ${ d_1 \gg \lambda_L (T_1) }$ is taken, which yields
\begin{equation}
\label{eqn:S10}
t_d (T_1, T_2) \approx \lambda_L (T_1) + \lambda_L (T_2) \tanh \left[ \frac{d_2}{2 \lambda_L (T_2)} \right] + d .
\end{equation}

Moreover, the Ambegaokar-Baratoff relation holds for the Josephson critical current
\begin{equation}
\begin{aligned}
\label{eqn:S11}
I_c (T_1, T_2) &= \frac{1}{2 e R} \left\vert \int_{- \infty}^{\infty}  \left\lbrace \mathcal{F} (\varepsilon, T_1) \mathrm{Re} \left[ \mathfrak{F} (\varepsilon, T_1) \right] \mathrm{Im} \left[ \mathfrak{F} (\varepsilon, T_2) \right] \right. \right. \\ & \left. \left. + \mathcal{F} (\varepsilon, T_2) \mathrm{Re} \left[ \mathfrak{F} (\varepsilon, T_2) \right] \mathrm{Im} \left[ \mathfrak{F} (\varepsilon, T_1) \right] \right\rbrace d \varepsilon \vphantom{\int_{- \infty}^{+ \infty} d \varepsilon} \right\vert ,
\end{aligned}
\end{equation}
in which ${ R = R_a / A }$ is the normal resistance, and ${ \mathfrak{F} (\varepsilon, T) = \frac{ \Delta (T) }{ \sqrt{ \left( \varepsilon + i \gamma \right)^2 - \Delta(T)^2 } }}$ is the anomalous Green's function~\cite{Guarcello_2019}. As a result, one gets the Josephson penetration depth ${ \lambda_J (T_1, T_2) = \sqrt{ \Phi_0 / \left[ 2 \pi \mu_0 t_d (T_1, T_2) J_c (T_1, T_2) \right]} }$, the normalized junction length ${ \mathcal{L} (T_1, T_2) = L / \lambda_J (T_1, T_2) }$, the Josephson plasma frequency ${ \omega_p (T_1, T_2) = \sqrt{ 2 \pi J_c (T_1, T_2) / \left( \Phi_0 C \right) } }$, the dissipation coefficient ${ \alpha (T_1, T_2) = 1 / \left[ \omega_p (T_1, T_2) R_a C \right] }$, and the noise strength ${ \Gamma (T_1, T_2) = 2 \pi \mathcal{L} (T_1, T_2) k_B T_1 / \left[ \Phi_0 I_c (T_1, T_2) \right] }$. Regarding the latter quantity, the explicit ${ T_1 }$ in the numerator indicates that the `worst-case' noise scenario is being accounted for.

Using the approximation ${ T_2 = T_b = 4.2 \; \mathrm{K} }$ (since ${ T_2 }$'s variations discussed in the paper are negligible for the present purpose), Fig.~\ref{fig:S1} displays ${ I_c }$ in units of ${ I_{c, 2} = I_c (T_2, T_2) }$~[panel~(a)], ${ \lambda_J }$~[panel~(b)], ${ \mathcal{L} }$~[panel~(c)], ${ \omega_p }$~[panel~(d)], ${ \alpha }$~[panel~(e)], and ${  \Gamma }$~[panel~(f)] versus the temperature ${ T_1 }$. The orange squares denote the parameter values at ${ T_1 = 7 \; \mathrm{K} }$, i.e., those employed throughout the work. Note also that, in addition to those listed above, the value ${ \lambda_L^0 = 80 \; \mathrm{nm} }$ is assumed here.

\begin{acknowledgments}

The authors acknowledge the support of the Italian Ministry of University and Research (MUR).

\end{acknowledgments}

\interlinepenalty=10000


%

\end{document}